%% file: isit11_final_03.tex
\documentclass[10pt,conference,letter]{IEEEtran}

\usepackage{amsfonts,amsmath,amssymb}
\usepackage{latexsym} 
\usepackage{mathrsfs}
\usepackage{graphicx}
\usepackage{indentfirst}
\usepackage{color}

\newtheorem{thm}{Theorem} 
\newtheorem{lem}[thm]{Lemma}
\newtheorem{co}[thm]{Corollary} 
\newtheorem{prop}[thm]{Proposition}

\newtheorem{ex}[thm]{Example}
\newtheorem{defi}[thm]{Definition}

\newcommand{\F}{\mathbb{F}}

\newcommand{\N}{\mathbb{N}}

\newcommand{\Gr}[3]{\mathcal{G}_{#1}(#2,#3)}

\newcommand{\Cvs}{\mathcal{C}}

\newcommand{\Vvs}{\mathcal{V}}
\newcommand{\Uvs}{\mathcal{U}}

\newcommand{\mat}[1]{\left(\begin{matrix}#1\end{matrix} \right)}
\newcommand{\rs}{\mathrm{rowsp}}

\newcommand{\Gg}{\mathfrak{S}}

\newcommand{\diag}{\mathrm{diag}}
\newcommand{\ord}{\mathrm{ord}}

\IEEEoverridecommandlockouts

\title{On conjugacy classes of subgroups of the general linear group
  and cyclic orbit codes}

\author{ \IEEEauthorblockN{ Felice Manganiello, Anna--Lena Trautmann
    and Joachim Rosenthal}
  \IEEEauthorblockA{
  Institute of Mathematics\\
  University of Zurich\\
  Winterthurerstrasse 190\\
  CH-8057 Zurich,  Switzerland\\
  \texttt{www.math.uzh.ch/aa} \thanks{The authors were
    partially supported by Swiss National Science Foundation under
    Grant no.\ 126948.}}}

\date{}

\begin{document}

\maketitle

\begin{abstract}
  Orbit codes are a family of codes applicable for communications on a
  random linear network coding channel. The paper focuses on the
  classification of these codes. We start by classifying the
  conjugacy classes of cyclic subgroups of the general linear
  group. As a result, we are able to focus the study of cyclic orbit
  codes to a restricted family of them.
\end{abstract}

\section*{Introduction}

The interest on constructions of codes for random linear network
coding arises with the paper \cite{ko08}. This paper introduces the
notion of a code as a subset of $\mathcal{P}(\Vvs)$, that is the set
of all subspaces of a vector space over a finite field $\F_q$. This
set is equipped with a metric, suitable for the model of communication
introduced, called subspace distance, defined as follows: for every
$\Uvs_1,\Uvs_2\in \mathcal{P}(\Vvs)$,
\[d(\Uvs_1,\Uvs_2)=\dim(\Uvs_1)+\dim(\Uvs_2)-2\dim(\Uvs_1\cap
\Uvs_2).\] The set of all subspaces of dimension $k$ is called the
Grassmannian and denoted by $\Gr{\F_q}{k}{n}$.

Some effort has been done in the direction of constructing codes for
random linear network coding in the last few years. Some results can
be found in \cite{ko08,ma08p,ko08p,et09,sk10,tr10p}.

In order to introduce orbit codes, we first recall the notion of the
right action of the group $GL_n(\F_q)$ of the invertible matrices on
the Grassmannian.

\begin{defi}
  Let $\Uvs \in \Gr{\F_q}{k}{n}$ and $U\in \F_q^{k\times n}$ a matrix
  such that $\Uvs:=\rs(U)$. We define the following operation
  \[\Uvs A:=\rs(UA).\]
  As a consequence we obtain the following right action of
  $GL_n(\F_q)$ on $\Gr{\F_q}{k}{n}$
  \begin{eqnarray*}
    \Gr{\F_q}{k}{n}\times  GL_n(\F_q) & \rightarrow & \Gr{\F_q}{k}{n}\\
    (\Uvs,A)& \mapsto & \Uvs A.
  \end{eqnarray*}
\end{defi}

The action just defined on $\Gr{\F_q}{k}{n}$ is independent of the
choice of the representation matrix $U\in \F_q^{k\times n}$ it is
distance preserving. For more information the reader is referred to
\cite{tr10p}.
 
Orbit codes are a certain class of constant
dimension codes.

\begin{defi}[\cite{tr10p}]
  Let $\Uvs\in \Gr{\F_q}{k}{n}$ and $\Gg<GL_n(\F_q)$ a subgroup. Then
  \[\Cvs = \{\Uvs A\mid A\in \Gg\}\]
  is called orbit code. An orbit code is called cyclic if there exists
  a subgroup defining it that is cyclic. 
\end{defi}

In \cite{tr10p} the authors show that orbit codes satisfy properties
that are similar to the ones of linear codes for classical coding
theory. Moreover, some already known
constructions, such as the ones contained in \cite{ko08} and
\cite{ma08p}, are actually orbit codes.

This paper focuses on the classification of
orbit codes. In order to do so, we are going to give a classification
of the conjugacy classes of subgroups of $GL_n(\F_q)$. 

The paper is structured as follows. The first section is dedicated to
the classification of subgroups of $GL_n(\F_q)$. More in detail, we
are able to characterize the properties of a unique representative
for the conjugacy classes of cyclic subgroups of $GL_n(\F_q)$. The
result is contained in Theorem \ref{t:rep_grp}. With some examples we
also show that the classification as it is cannot be
extended to arbitrary subgroups. In the second section we apply these
results to cyclic orbit codes. The main result is that we can focus on
the study of  cyclic orbit codes defined by a cyclic group
generated by a matrix in rational canonical form. Moreover we study
the construction of codes in this case and relate them to
completely reducible cyclic orbit codes. At last we give some
conclusions.

\section{Characterization of cyclic subgroups of $GL_n(\F_q)$}\label{s:1}

In this section we investigate the cyclic subgroups of
$GL_n(\F_q)$. The goal is to characterize them in a way that is
suitable for the construction of orbit codes. More specifically we are
interested in answering the question about when two cyclic groups are
conjugate to each other.

Consider $GL_n(\F_q)$ and the following equivalence relation on it:
Given $A,B\in GL_n(\F_q)$ then
\[A\sim_c B \quad \iff \quad \exists L\in GL_n(\F_q): \ A=L^{-1}BL.\]
A natural choice of representatives of the classes of $
GL_n(\F_q)/\sim_c$ is given by the \emph{rational canonical
  form}. Rational canonical forms are based on companion matrices,
whose definition is as follows.

\begin{defi}
  Let $p=\sum_{i=0}^sp_ix^i\in \F_q[x]$ be a monic polynomial. Its
  companion matrix is the matrix 
  \[M_p:=\mat{0&1&0 &\cdots &0\\0 & 0 & 1 & &0\\\vdots & & & \ddots
    &\vdots\\ 0&0&0& & 1 \\ -p_0& -p_1 & -p_2 & \cdots & -p_{s-1}}\in
  \F_q^{s\times s}.\]
\end{defi}

The following theorem states the existence and uniqueness of a
rational canonical form.

\begin{thm}[{\cite[Chapter~6.7]{he75}}]
  Let $A\in GL_n(\F_q)$. Then there exists a matrix $L\in GL_n(\F_q)$
  such that 
  \begin{align}
    \label{e:rcf} \nonumber
    L^{-1}AL = \diag(&M_{p_1^{e_{11}}},\dots, M_{p_1^{e_{1r_1}}},\\
    &\dots,M_{p_m^{e_{m1}}}, \dots, M_{p_m^{e_{mr_m}}})
  \end{align}
  is a block diagonal matrix where $p_i\in \F_q[x]$ are irreducible
  polynomials, $e_{ij}\in \N$ are  such that $e_{i1}\geq \dots \geq
  e_{ir_i}$, $\chi_A=\prod_{i,j}p_i^{e_{ij}}$ and $\mu_A=\prod_i
  p_i^{e_{i1}}$ represent respectively the characteristic and the
  minimal polynomials of $A$ and $M_{p_i^{e_{ij}}}$ denotes the
  companion matrix of the polynomial $p^{e_{ij}}$. Moreover, the
  matrix \eqref{e:rcf} is unique for any choice of $A\in GL_n(\F_q)$.
\end{thm}

\begin{defi}\label{d:rcf}
  Let $A\in GL_n(\F_q)$. The matrix \eqref{e:rcf} is called rational
  canonical form of $A$ and the polynomials
  $p_1^{e_{11}},\dots,p_1^{e_{1r_1}},\dots,p_m^{e_{m1}},\dots,p_m^{e_{mr_m}}
  \in \F_q[x]$ are its elementary divisors.
\end{defi}


The following lemma motivates why rational canonical forms are a good
choice of representatives for the classes of $GL_n(\F_q)/\sim_c$.

\begin{lem}\label{l:conj_mat}
  Let $A,B\in GL_n(\F_q)$. Then the following statements are
  equivalent:
  \begin{enumerate}
  \item $A\sim_c B$, and
  \item $A$ and $B$ have the same rational canonical form. 
  \end{enumerate}
\end{lem}

This lemma is well-known and is a direct consequence of the uniqueness
of the rational canonical form.

Now we want to extend the previous characterization to
subgroups of $GL_n(\F_q)$.

Consider the set of all subgroups of $GL_n(\F_q)$
\[\mathbf{G}:=\{\Gg \mid \Gg < GL_n(\F_q)\}\]
and the following equivalence relation on it. Given
$\Gg_1,\Gg_2\in \mathbf{G}$ then
\[\Gg_1\sim_c \Gg_2 \quad \iff \quad \exists L\in GL_n(\F_q): \
\Gg_1=L^{-1}\Gg_2L.\]

The following theorem extends the arguments of Lemma \ref{l:conj_mat}
to the case of cyclic subgroups.

\begin{thm}\label{t:conj_grp}
  Let $A,B\in GL_n(\F_q)$ and $\Gg_A=\langle A\rangle, \Gg_B=\langle
  B\rangle< GL_n(\F_q)$ be the two cyclic groups generated by
  them. Then, $\Gg_A\sim_c \Gg_B$ if and only if $|\Gg_A|=|\Gg_B|$ and
  there exists an $i\in \N$ with $\gcd(i,|\Gg_B|)=1$ such that
  $A\sim_cB^i$.
\end{thm}

\begin{IEEEproof}
  \begin{description}
  \item[\fbox {$\Rightarrow$}] Since $\Gg_A\sim_c \Gg_B$, it follows
    that there exists an $L\in GL_n(\F_q)$ such that
    $\Gg_A=L^{-1}\Gg_BL$, implying that the two groups have the same
    order. Moreover, it follows that the group homomorphism
    \begin{eqnarray*}
      \varphi:\Gg_A & \rightarrow & GL_n(\F_q) \\
      A^i & \mapsto & LA^iL^{-1}
    \end{eqnarray*}
    is an isomorphism if restricted to the image of $\varphi$. As a
    consequence, the generator $A$ of $\Gg_A$ is mapped to a
    generator of $L\Gg_AL^{-1}=\Gg_B$, i.e., an element of $\{B^i\mid
    \gcd(i,|\Gg_B|)=1\}$. Then, there exists an $i\in \N$ with
    $\gcd(i,|\Gg_B|)=1$ such that $LAL^{-1}=B^i$, which implies that
    $A\sim_c B^i$.
  \item[\fbox {$\Leftarrow$}] 
    From the hypothesis we know that $\langle B^i \rangle =\Gg_B$ and
    that there exists $L\in GL_n(\F_q)$ such that $A=L^{-1}B^iL$. The
    statement follows as a consequence.
  \end{description}
\end{IEEEproof}

We introduce the following definition.

\begin{defi}[{\cite[Definition~3.2]{li94}}]
  Let $p\in \F_q[x]$ be a nonzero polynomial. If $p(0)\neq 0$, then
  the least integer $e\in \N$ such that $p$ divides $x^e-1$ is called
  the order of $p$.
\end{defi}

The definition is generalizable to any $p\in \F_q[x]$ but it is not
interesting for the purpose of this paper since we will only consider
irreducible polynomials.

In order to give unique representatives for the classes of cyclic
groups contained in $\mathbf{G}/\sim_c$ we need the following lemma.

\begin{lem}\label{l:elem_div}
  Let $A\in GL_n(\F_q)$, $p_{A,1}^{e_{A,1}},\dots,p_{A,m}^{e_{A,m}}
  \in \F_q[x]$ its elementary divisors, where $p_{A,j}$ for $j\in
  \{1,\dots,m\}$ are not necessarily distinct, and $\Gg_A< GL_n(\F_q)$
  the cyclic group generated by $A$.  Then, for every $i\in \N$ with
  $\gcd(i,|\Gg_A|)=1$, the elementary divisors of $A^i$ are exactly
  $m$ many. If we denote them by $p_{A^i,1}^{e_{A^i,1}},\dots,
  p_{A^i,m}^{e_{A^i,m}}\in \F_q[x]$, then, up to reordering, the order
  of $p_{A,j}$ is the same as the one of $p_{A^i,j}$ and
  $e_{A,j}=e_{A^i,j}$ for $j=1,\dots,m$.
\end{lem}

\begin{IEEEproof}
  First we prove the case where the elementary divisor is unique. At
  the end of the proof we will give the main remark that implies the
  generalized statement.

  Let $p_A^{e_A}\in \F_q[x]$ be the elementary divisor of a matrix
  $A\in GL_n(\F_q)$ and $k:=n/e_A$. Let $\F_{q^k}:=\F_q[x]/(p_A)$ be
  the splitting field of the polynomial $p_A$ and $\mu\in \F_{q^k}$ a
  primitive element of it. There exists a $j\in \N$ such that
  $p_A=\prod_{u=0}^{k-1}(x-\mu^{jq^u})$. Since $p_A^{e_A}$ is the
  unique elementary divisor of the matrix $A$, it corresponds to the
  characteristic and the minimal polynomial of $A$. As a consequence
  we obtain that the Jordan normal form of $A$ over $\F_{q^k}$ is
  \[J_A=\diag\left(J_{A,\mu^j}^{e_a},\dots,J_{A,\mu^{jq^{k-1}}}^{e_a}\right)\]
  where $J_{A,\mu^{jq^u}}^{e_a}\in GL_{e_A}(\F_{q^k})$ is a unique
  Jordan block with diagonal entries $\mu^{jq^u}$ for $u=0,\dots,k-1$.

  By the Jordan normal form of $A$ it follows that for every $i\in \N$
  the characteristic polynomial of $A^i$ is
  $p_{A^i}=(\prod_{u=0}^{k-1}x-\mu^{ijq^u})^{e_A}$. Let us now focus
  on the $i$'s such that $\gcd(i,|\Gg_A|)=1$. $A^i$ is then a
  generator of $\Gg_A$, i.e., $p_{A^i}\in \F_q[x]$ is a monic
  irreducible polynomial whose order is the same as the one of $p_A$.

  In order to conclude that $p_{A^i}^{e_A}$ is the elementary divisor
  of $A^i$ we consider its rational canonical form. Assume that the
  elementary divisors of $A^i$ were more than one. Without loss of
  generality we can consider them to be two, i.e., $p_{A^i}^{e_{A,1}}$
  and $p_{A^i}^{e_{A,2}}$. This means that its rational canonical form
  is $\mathrm{RCF}(A^i)=
  \diag(M_{p_{A^i}^{e_{A,1}}},M_{p_{A^i}^{e_{A,2}}})$ where we use the
  operator $\mathrm{RCF}$ as an abbreviation for rational canonical
  form and $e_A=e_{A,1}+e_{A,2}$. For any $j\in \N$ we obtain that the
  matrix $\mathrm{RCF}((\mathrm{RCF}(A^i))^j)$ is a block diagonal
  matrix with at least two blocks. Let $j\in \N$ such that $ij\equiv 1
  \pmod{|\Gg_A|}$ and $L\in GL_n(\F_q)$ be a matrix such that
  $\mathrm{RCF}(A^i)=L^{-1}A^iL$, then
  \[(\mathrm{RCF}(A^i))^j=(L^{-1}A^iL)^j=L^{-1}AL\sim_c A\]
  implying that
  \[\mathrm{RCF}(A)=\mathrm{RCF}((\mathrm{RCF}(A^i))^j)\] 
  This leads to a contradiction since $\mathrm{RCF}(A)=M_{p_A^{e_A}}$
  has only one block. We conclude that $p_{A^i}^{e_A}$ is the
  elementary divisor of $A^i$.

  The only difference in the case where $m>1$ consists in the choice
  of the splitting field. Given
  $p_{A,1}^{e_{A,1}},\dots,p_{A,m}^{e_{A,m}} \in \F_q[x]$ the
  elementary divisors of $A$ and $p_{A,l_1},\dots p_{A,l_r}$ with
  $l_1,\dots l_r \in \{1,\dots,m\}$ the maximal choice distinct
  polynomials from the elementary divisors, the splitting field on
  which the proof is based is $\F_q[x]/(\prod_{t=1}^r p_{A,l_t})$.
\end{IEEEproof}

We are now ready to characterize cyclic subgroups of $GL_n(\F_q)$ via
the equivalence relation $\sim_c$ based only on their elementary
divisors.

\begin{thm}\label{t:rep_grp}
  Let $A,B\in GL_n(\F_q)$ and $\Gg_A,\Gg_B\in \mathbf{G}$ the cyclic
  subgroups generated by them. Then, $\Gg_A\sim_c \Gg_B$ if and only if
  the following conditions hold:
  \begin{enumerate}
  \item $A$ and $B$ have the same number of elementary divisors, and
  \item if $p_{A,1}^{e_{A,1}},\dots,p_{A,m}^{e_{A,m}} \in \F_q[x]$ and
    $p_{B,1}^{e_{B,1}},\dots,p_{B,m}^{e_{B,m}} \in \F_q[x]$ are the
    elementary divisors of respectively $A$ and $B$, then, up to
    a reordering argument, the orders of $p_{A,j}$ and $p_{B,j}$ are
    the same and $e_{A,j}=e_{B,j}$ for $j=1,\dots,m$.
  \end{enumerate}
\end{thm}

\begin{IEEEproof}
  \begin{description}
  \item[\fbox {$\Rightarrow$}] By Theorem \ref{t:conj_grp}, there
    exists a power $i\in \N$ with $\gcd(i,|\Gg_A|)=1$ such that
    $A\sim_c B^i$, i.e., they have the same elementary divisors. The
    statement follows with Lemma \ref{l:elem_div}.
  \item[\fbox {$\Leftarrow$}] Let $p_{B,l_1},\dots p_{B,l_r}\in
    \F_q[x]$ with $l_1,\dots l_r \in \{1,\dots,m\}$ be the maximal
    choice of pairwise coprime polynomials from the elementary
    divisors of $B$, $\F$ the splitting field of
    $\prod_{t=1}^rp_{B,l_t}$ and $\mu\in \F$ a primitive element of
    it. Consider the notation $k_j:= \deg p_{B,l_j}$ for $j=1,\dots,
    r$. Then, there exist $i_{B,1},\dots,i_{B,r}\in \N$ such that
    $p_{B,l_j}=\prod_{u=0}^{k_j-1}(x-\mu^{i_{B,j}q^u})$ for
    $j=1,\dots,r$. The same holds for the matrix $A$, i.e., there
    exist $i_{A,1},\dots,i_{A,r}\in \N$ such that
    $p_{A,l_j}=\prod_{u=0}^{k_j-1}(x-\mu^{i_{A,j}q^u})$ for
    $j=1,\dots,r$. By the condition on the orders, there exists a
    unique $i\in \N$ such that $i_{A,j}\equiv i\cdot i_{B,j}
    \pmod{\ord(p_{B,l_j})}$ for $j=1,\dots,r$. It follows that the
    elementary divisors of $B^i$ and the ones of $A$ are the same,
    i.e., $A\sim_c B^i$.
  \end{description}
\end{IEEEproof}

The theorem states that we can uniquely represent the classes of
cyclic subgroups in $\mathbf{G}/\sim_c$ by considering the cyclic subgroups
generated by a rational canonical form based on the choice of a
sequence of polynomials of the type $p_1^{e_1},\dots,p_m^{e_m}\in
\F_q[x]$ where the polynomials $p_1,\dots,p_m$ are irreducible and
$\sum_{j=1}^me_j\cdot \deg(p_j)=n$. Moreover, what matters in the
choice of the polynomials $p_j$'s is only their degrees and orders.

Trivially, the following holds for the cardinality of a cyclic
group.

\begin{co}
  Let $\Gg_A=\langle A \rangle < GL_n(\F_q)$. Then the order of
  $\Gg_A$ is the least common multiple of the orders of the elementary
  divisors $p_1^{e_1},\dots,p_m^{e_m}\in \F_q[x]$ of the matrix $A$.
\end{co}

To conclude the section we are going to give an example explaining why
a straight forward generalization of Theorem \ref{t:rep_grp}  to any subgroup of $GL_n(\F_q)$ does not work.

\begin{ex} {\ }
  \begin{enumerate}
  \item Consider the following matrix over $\F_2$:
    \[A=\mat{0 & 1 & 0 \\ 0 & 0 & 1 \\ 1 & 1 & 0}.\] Although the
    elementary divisor of $A$ and the one of its transpose $A^t$ is
    the same, the groups $\Gg_A=\langle A\rangle = \langle A,A\rangle$
    and $GL_3(\F_2)=\langle A,A^t\rangle$ are not conjugate.
  \item Let $\F_{4}=\F_2[x]/(x^2+x+1)$ and $\mu\in \F_4$ a primitive
    element. Consider the following matrices over $\F_4$:
    \begin{align*}A=\mat{0 & 1 & 0 \\ 0 & 0 & 1 \\ 1 & 1 & 0}, \
      &B_1=\mat{0 & 1 & 0 \\ 0 & 0 & 1 \\ 1 & 0 & 1},\\ \mbox{ and }
      &B_2=\mat{\mu+1 & 1 & \mu \\ \mu & \mu & \mu+1 \\ 0 & 1 &
        0}.\end{align*} Although $B_1\sim_c B_2$, i.e., they have the
    same unique elementary divisor, it holds that $|\langle
    A,B_1\rangle|\neq |\langle A,B_2\rangle|$, meaning that the two
    groups are not conjugate.
  \end{enumerate}
\end{ex}

\section{Conjugate groups and  cyclic orbit codes}

We now apply the results from the previous section to the
characterization of cyclic codes.

\begin{defi}
  Let $\Gg_1,\Gg_2<GL_n(\F_q)$ and $\Cvs_1:=\{\Uvs_1A\mid A\in
  \Gg_1\},\Cvs_2:=\{\Uvs_2A\mid A\in \Gg_2\}\subseteq \Gr{\F_q}{k}{n}$
  be two orbit codes. We say that $\Cvs_1$ and $\Cvs_2$ are conjugate
  or simply $\Cvs_1\sim_c \Cvs_2$ if there exists a matrix $L\in
  GL_n(\F_q)$ such that 
  \[\Uvs_2=\Uvs_1L \mbox{ and } \Gg_2=L^{-1}\Gg_1L,\]
  i.e., $\Cvs_2=\{\Uvs_1AL\mid A\in \Gg_1\}=\{\Uvs_1L(L^{-1}AL)\mid
  A\in \Gg_1\}$.
\end{defi}

In order to further study properties of orbit codes, we need to
introduce the notion of distance distribution for orbit codes. Due to 
\cite{tr10p}, we are able to adapt the definition of
weight enumerator from classical coding theory to orbit codes. But
first we recall some facts from \cite{tr10p}.

\begin{defi}[{\cite[Definition~3]{tr10p}}]
  Let $\Uvs\in \Gr{\F_q}{k}{n}$. Then the stabilizer group of $\Uvs$
  is defined as 
  \[Stab(\Uvs):=\{A\in GL_n(\F_q)\mid \Uvs A=\Uvs\}<GL_n(\F_q).\]
\end{defi}

The following proposition is important in order to define the distance
distribution.

\begin{prop}[{\cite[Proposition~8]{tr10p}}]
  Let $\Cvs=\{\Uvs A\mid A\in \Gg<GL_n(\F_q)\}$ be an orbit code. Then
  it holds that 
  \[|\Cvs|=\frac{|\Gg|}{|\Gg\cap Stab(\Uvs)|}\]
  and 
  \[d(\Cvs)=\min_{A\in \Gg\setminus Stab(\Uvs)}d(\Uvs,\Uvs A).\]
\end{prop}

\begin{defi}
  Let $\Cvs=\{\Uvs A\mid A \in \Gg< GL_n(\F_q)\}\subseteq
  \Gr{\F_q}{k}{n}$ be an orbit code. The distance distribution of
  $\Cvs$ is the tuple $(D_0,\dots,D_k)\in \N^{k+1}$ such that
  \[D_i:=\frac{|\{A\in \Gg\mid d(\Uvs,\Uvs A)=2i\}|}{|\Gg \cap Stab(\Uvs)|}.\]
\end{defi}

As a consequence we obtain that $D_0=1$ and
$\sum_{i=0}^kD_i=|\Cvs|$. We are able to state the following theorem
that characterizes conjugate orbit codes and that is a generalization
of Theorem 9 from \cite{tr11}.

\begin{thm}\label{t:conj_orb}
  The binary relation $\sim_c$ on orbit codes is an equivalence
  relation. Moreover, let $\Cvs_1,\Cvs_2$ be two orbit codes such that
  $\Cvs_1\sim_c \Cvs_2$, then $|\Cvs_1|=|\Cvs_2|$ and they have the
  same distance distribution. 
\end{thm}

\begin{IEEEproof}
  The fact that $\sim_c$ is an equivalence relation on orbit codes
  is a consequence of Theorem \ref{t:conj_grp}.
  
  Let $\Cvs_1:=\{\Uvs A\mid A\in \Gg<GL_n(\F_q)\}$ and $L\in
  GL_n(\F_q)$ such that $\Cvs_2=\{\Uvs AL\mid A\in \Gg\}$. The same
  cardinality is consequence of the fact that given $A,B\in \Gg$ then
  \[\Uvs AL=\Uvs BL \iff \Uvs A=\Uvs B.\]
  The same distance distribution follows from the distance preserving
  property of the $GL_n(\F_q)$ action on $\Gr{\F_q}{k}{n}$, i.e.,
  $d(\Uvs L,\Uvs AL)=d(\Uvs,\Uvs A)$.
\end{IEEEproof}

The importance of this last theorem is that two conjugate orbit codes
are not distinguishable from the point of view of cardinality and
distance distribution.  Theorem \ref{t:rep_grp} translates as follows
in the language of orbit codes.

\begin{co}
  Every cyclic orbit code is conjugate to a cyclic orbit code defined
  by a cyclic group generated by a matrix in rational canonical form.
\end{co}

This fact gives us the opportunity to consider only cyclic orbit codes
out of matrices in rational canonical form for the study of codes with
good parameters.

We are now interested in these orbits codes.

\begin{thm}\label{t:codes}
  Let $M:=\diag(M_{p_1^{e_1}},\dots,M_{p_t^{e_t}})\in GL_n(\F_q)$ a
  matrix such that $p_i\in \F_q[x]$ are monic irreducible polynomials
  and $d_i:=\deg(p_i^{e_i})$ for $i=1,\dots,t$. Let
  $\Uvs=\rs(U_1,\dots,U_t)\in \Gr{\F_q}{k}{n}$ with $U_i\in
  \F_q^{k\times d_i}$ and where $(U_1,\dots,U_t)$ is in row reduced
  echelon form. For any $i\in \{1,\dots,t\}$, let $\bar{U}_i$ be a
  submatrix of $U_i$ as depicted in Figure \ref{f:rre}.
  
  \begin{figure}[h]
    \centering
    \input{rre.pstex_t}
      \caption{The matrix $U$ in row reduced echelon form.}\label{f:rre}
  \end{figure}
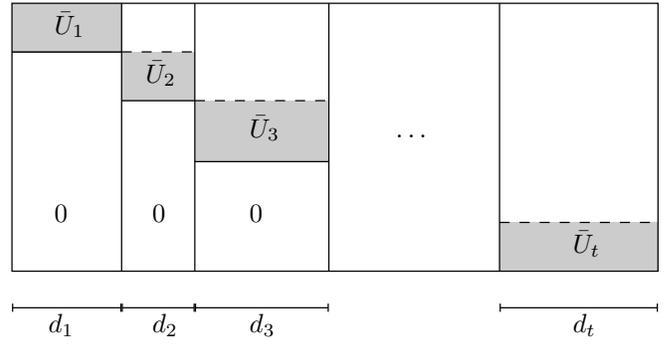

  If $\Cvs:=\{\Uvs M^i\mid i\in \N\}$ and
  $\Cvs_i:=\{\rs(\bar{U}_i)M_{p_i^{e_i}}^j\mid j\in \N\}$, then
  \begin{eqnarray}\label{e:modular_dmin}
    d(\Cvs)\!\geq\! 2k\!-\!2\!\sum_{i=1}^t \!\max_{j\in \N}
  \dim\!\left(\!\rs\left(\bar{U}_i\right) \!\cap\!
    \rs\left(\!\bar{U}_iM_{p_i^{e_i}}^j\!\right)\!\right)\!,\!\!\!
  \end{eqnarray}
  and $|\Cvs|:=\mathrm{lcm}(|\Cvs_i|,\dots,|\Cvs_t|)$.

\end{thm}

\begin{IEEEproof}
  Consider the following projections 
  \[
  \begin{array}{rccc}
    \pi_i:&\F_q^n & \longrightarrow &\F_q^{d_i}\\
    &(v_1,\dots,v_n) & \longmapsto & (v_{l_{i-1}+1},\dots,v_{l_i})
  \end{array}
  \]
  where $l_i=\sum_{j=1}^id_i$ for $i=1,\dots,t$. Since
  $(U_1,\dots,U_t)$ has full rank and is in row reduced echelon form,
  the matrices $\bar{U}_i$ have full rank. Let $\bar{\Uvs}_i\subset
  \F_q^n$ be the space spanned by the rows of $(U_1,\dots,U_t)$
  indexed by the rows corresponding to $\bar{U}_i$. Since $\bar{U}_i$
  has full rank it follows that $\pi_i\vert_{\bar{\Uvs}_i}$ is
  injective for $i=1,\dots,t$. As a consequence we obtain that for any
  $i=1,\dots,t$, if we
  define $m_i\in \N$ such that 
  \[\dim \left(\bar{\Uvs}_i\cap
  \bar{\Uvs}_iM_{p_i^{e_i}}^{m_i}\right)\geq \dim \left(\bar{\Uvs}_i\cap
  \bar{\Uvs}_iM_{p_i^{e_i}}^j\right), \quad \forall j\in \N\] 
  and $\Vvs_i:= \bar{\Uvs}_i\cap \bar{\Uvs}_iM_{p_i^{e_i}}^{m_i}$, 
  then
  \[\pi_i(\Vvs_i)\subseteq
  \rs(\bar{U}_i)\cap\rs(\bar{U}_iM_{p_i^{e_i}}^{m_i}).\]
  It follows that 
  \[\dim(\Vvs_i)\leq \max_{j\in
    \N}\dim(\rs(\bar{U}_i)\cap\rs(\bar{U}_iM_{p_i^{e_i}}^{j})).\]
  Since $\Uvs=\oplus_{i=1}^t\bar{\Uvs}_i$ we conclude that 
  \begin{align*}
    d(\Cvs)&=2k-2\max_{j\in \N}\dim (\Uvs\cap\Uvs M^j)\\
    &\geq 2k-2\sum_{i=1}^t \max_{j\in \N}
    \dim\left(\rs\left(\bar{U}_i\right) \cap
      \rs\left(\bar{U}_iM_{p_i^{e_i}}^j\right)\right)
  \end{align*}

  The cardinality of $\Cvs$ is a direct consequence of the fact that
  \[\diag(M_{p_1^{e_1}},\dots,M_{p_t^{e_t}})^i=
  \diag(M_{p_1^{e_1}}^i,\dots,M_{p_t^{e_t}}^i)\] 
  and of the minimality of the least common multiple.
\end{IEEEproof}

It is possible to find examples for which the lower bound given by
\eqref{e:modular_dmin} is attained. The following lemmas depict
these examples.

\begin{lem}
  Let $M:=\diag(M_{p_1^{e_1}},\dots,M_{p_t^{e_t}})\in GL_n(\F_q)$ a
  matrix such that $p_i\in \F_q[x]$ are monic irreducible polynomials
  and $d_i:=\deg(p_i^{e_i})$ for $i=1,\dots,t$. Let $k\leq d_i$ for
  $i=1,\dots,t$ and $\Uvs:=\rs (U_1,\dots,U_t)\in \Gr{\F_q}{k}{n}$
  where $U_i\in \F_q^{k\times d_i}$ are matrices having full rank
  for $i=1,\dots,t$.  If we define $\Cvs:=\{\Uvs M^i\mid i\in \N\}$ and
  $\Cvs_i:=\{\rs(U_i)M_{p_i^{e_i}}^j\mid j\in \N\}$ and it holds
  $\gcd(|\Cvs_i|,|\Cvs_j|)=1$ for all $i\neq j$, then
  \begin{eqnarray*}
    d(\Cvs)= \min_{i\in \{1,\dots, t\}}d(\Cvs_i).
  \end{eqnarray*}
\end{lem}

\begin{IEEEproof}
  We only need to show that there exists a codeword of $\Cvs$ that
  satisfies this minimum. Up to a permutation of $\{1,\dots,t\}$ we
  can consider that the code $\Cvs_1$ is satisfying the minimum
  distance. Let $g_1\in \N$ be such that
  $d(\rs(U_1),\rs(U_1)M_{p_1^{e_1}}^{g_1})=d(\Cvs_1)$. Since the
  cardinalities of the codes $\Cvs_i$ are pairwise coprime, it follows
  that there exists $g\in \N$ such that
  \[g \equiv g_1 \pmod{|\Cvs_1|}\quad \mbox{and}\quad g\equiv 0
  \pmod{|\Cvs_j|}\] 
  for $j=2,\dots,m$. We obtain that
  \begin{eqnarray*}
    d(\Uvs,\Uvs M^g)&=&d(\Uvs,\Uvs\diag(M_{p_1^{e_1}}^{g_1},I,\dots,I))\\
    &=&d(\rs(U_1),\rs(U_1)M_{p_1^{e_1}}^{g_1})=d(\Cvs_1)
  \end{eqnarray*}
\end{IEEEproof}

\begin{lem}
  Let $M:=\diag(M_{p_1^{e_1}},\dots,M_{p_t^{e_t}})\in GL_n(\F_q)$ such
  that $p_i\in \F_q[x]$ are monic irreducible polynomials and
  $d_i:=\deg(p_i^{e_i})$ for $i=1,\dots,t$. Let $k_i\leq d_i$,
  $\bar{U}_i\in \F_q^{k_i\times d_i}$ be matrices with full rank and
  $\Uvs:=\diag(\bar{U}_1,\dots,\bar{U}_t)\in \Gr{\F_q}{k}{n}$. If we
  define $\Cvs:=\{\Uvs M^i\mid i\in \N\}$ and
  $\Cvs_i:=\{\rs(\bar{U}_iM_{p_i^{e_i}})^j\mid j\in \N\}$ and it holds
  $\gcd(|\Cvs_i|,|\Cvs_j|)=1$ for all $i\neq j$, then
  \begin{eqnarray*}
    d(\Cvs)\!= \!2k-2\!\sum_{i=1}^t \max_{j\in \N}
  \dim\!\left(\rs\left(\bar{U}_i\right) \cap
    \rs\left(\!\bar{U}_iM_{p_i^{e_i}}^j\!\right)\right).
  \end{eqnarray*}
\end{lem}

\begin{IEEEproof}
  Also here we show a codeword of $\Cvs$ which satisfies the
  relation. Let $g_1,\dots,g_t\in \N$ be such that $\dim(\rs(\bar{U}_j)\cap
  \rs(\bar{U}_jM_{p_j^{e_j}}^{g_j})$ is maximal for $j=1,\dots,m$. Since the
  cardinalities of the codes are pairwise coprime, it follows that
  there exists a $g\in\N$ such that 
  \[g\equiv g_j \pmod{|\Cvs_j|}\] 
  for any $j=1,\dots,t$. Then,

  \begin{align*}
    d_{\min}(\Cvs)&= d(\Uvs,\Uvs\diag(M_{p_1^{e_1}},\dots,M_{p_m^{e_m}})^g)\\
    &=d(\Uvs,\Uvs\diag(M_{p_1^{e_1}}^{g_1},\dots,M_{p_m^{e_m}}^{g_m}))\\
    &\phantom{=2k-2\sum_{j=1}^m\dim(\rs(\bar{U}_j)\cap\rs(\bar{U}_jM_{p_j^{e_j}}^{g_j})).}
  \end{align*}
  \newpage
  \begin{align*}
    \phantom{d_{\min}(\Cvs)}&=2k-2\sum_{j=1}^m\dim(\rs(\bar{U}_j)\cap\rs(\bar{U}_jM_{p_j^{e_j}}^{g_j})).
  \end{align*}
\end{IEEEproof}

A matrix $M\in GL_n(\F_q)$ is called completely reducible if its
elementary divisors are all irreducible, i.e., from Definition
\ref{d:rcf} if $e_{i,j}=1$ for all $i,j$. One can use the theory of
irreducible cyclic orbit codes from \cite{tr11} to compute the minimum
distances of the block component codes in the extension field
representation and hence with Theorem \ref{t:codes} a lower bound for
the minimum distance of the whole code.

\section*{Conclusions}

Due to the characterization of conjugacy classes of cyclic subgroups of
$GL_n(\F_q)$, we were able to conclude that every cyclic orbit code is
conjugated to a cyclic orbit code defined by the cyclic group
generated by a matrix in rational canonical form. The research of
orbit codes with good parameters can then be restricted to this
subclass of cyclic orbit codes. 

The following step in this research direction is to completely
classify orbit codes. In order to do so we have to find a
characterization of the conjugacy classes of subgroups of $GL_n(\F_q)$
that possibly coincides with the one presented in Section \ref{s:1} if
restricted to cyclic subgroups of $GL_n(\F_q)$.

\bibliographystyle{IEEEtran}
\bibliography{../../huge,../../to_update_publications}
\end{document}

%% file: rre.pstex_t
\begin{picture}(0,0)%
\includegraphics{rre.pstex}%
\end{picture}%
\setlength{\unitlength}{4144sp}%
\begingroup\makeatletter\ifx\SetFigFont\undefined%
\gdef\SetFigFont#1#2#3#4#5{%
  \reset@font\fontsize{#1}{#2pt}%
  \fontfamily{#3}\fontseries{#4}\fontshape{#5}%
  \selectfont}%
\fi\endgroup%
\begin{picture}(3889,2047)(3094,-3771)
\put(4528,-3048){\makebox(0,0)[lb]{\smash{{\SetFigFont{10}{12.0}{\rmdefault}{\mddefault}{\updefault}{\color[rgb]{0,0,0}$0$}%
}}}}
\put(3361,-1918){\makebox(0,0)[lb]{\smash{{\SetFigFont{10}{12.0}{\rmdefault}{\mddefault}{\updefault}{\color[rgb]{0,0,0}$\bar{U}_1$}%
}}}}
\put(3908,-2210){\makebox(0,0)[lb]{\smash{{\SetFigFont{10}{12.0}{\rmdefault}{\mddefault}{\updefault}{\color[rgb]{0,0,0}$\bar{U}_2$}%
}}}}
\put(4528,-2538){\makebox(0,0)[lb]{\smash{{\SetFigFont{10}{12.0}{\rmdefault}{\mddefault}{\updefault}{\color[rgb]{0,0,0}$\bar{U}_3$}%
}}}}
\put(3361,-3048){\makebox(0,0)[lb]{\smash{{\SetFigFont{10}{12.0}{\rmdefault}{\mddefault}{\updefault}{\color[rgb]{0,0,0}$0$}%
}}}}
\put(3945,-3048){\makebox(0,0)[lb]{\smash{{\SetFigFont{10}{12.0}{\rmdefault}{\mddefault}{\updefault}{\color[rgb]{0,0,0}$0$}%
}}}}
\put(3325,-3705){\makebox(0,0)[lb]{\smash{{\SetFigFont{10}{12.0}{\rmdefault}{\mddefault}{\updefault}{\color[rgb]{0,0,0}$d_1$}%
}}}}
\put(3945,-3705){\makebox(0,0)[lb]{\smash{{\SetFigFont{10}{12.0}{\rmdefault}{\mddefault}{\updefault}{\color[rgb]{0,0,0}$d_2$}%
}}}}
\put(4528,-3705){\makebox(0,0)[lb]{\smash{{\SetFigFont{10}{12.0}{\rmdefault}{\mddefault}{\updefault}{\color[rgb]{0,0,0}$d_3$}%
}}}}
\put(6461,-3705){\makebox(0,0)[lb]{\smash{{\SetFigFont{10}{12.0}{\rmdefault}{\mddefault}{\updefault}{\color[rgb]{0,0,0}$d_t$}%
}}}}
\put(6461,-3231){\makebox(0,0)[lb]{\smash{{\SetFigFont{10}{12.0}{\rmdefault}{\mddefault}{\updefault}{\color[rgb]{0,0,0}$\bar{U}_t$}%
}}}}
\put(5403,-2538){\makebox(0,0)[lb]{\smash{{\SetFigFont{10}{12.0}{\rmdefault}{\mddefault}{\updefault}{\color[rgb]{0,0,0}$\dots$}%
}}}}
\end{picture}%